\documentclass{sigchi}
\usepackage{times}
\usepackage[utf8]{inputenc}

\title{Toward Finding Latent Cities \\ with Non-Negative Matrix Factorization}

\toappear{
\copyright 2018. Copyright for the individual papers remains with the authors. Copying permitted for private and academic purposes. \\
\emph{UISTDA '18},  March 11, 2018, Tokyo, Japan
}
\numberofauthors{3}
\author{
\alignauthor
Eduardo Graells-Garrido \\
\affaddr{Data Science Institute \\ Universidad del Desarrollo} \\
\affaddr{Telefonica R\&D} \\
\affaddr{Santiago, Chile} \\
\email{egraells@udd.cl}
\and
\alignauthor 
Diego Caro \\
\affaddr{Data Science Institute \\ Universidad del Desarrollo} \\
\affaddr{Telefonica R\&D} \\
\affaddr{Santiago, Chile} \\
\email{dcaro@udd.cl}
\and
\alignauthor
Denis Parra \\
\affaddr{Dept. of Computer Science} \\
\affaddr{Faculty of Engineering} \\
\affaddr{Pontificia Universidad Catolica} \\
\affaddr{Santiago, Chile} \\
\email{dparra@ing.puc.cl}
}

\usepackage{balance}
\usepackage{txfonts}
\usepackage{color}
\usepackage{booktabs}
\usepackage{textcomp}
\usepackage{microtype}
\usepackage{ccicons}
\usepackage{subcaption}

\providecommand{\tightlist}{%
  \setlength{\itemsep}{0pt}\setlength{\parskip}{0pt}}

\begin{document}

\maketitle

\begin{abstract}
In the last decade, digital footprints have been used to cluster population activity into functional areas of cities.
  However, a key aspect has been overlooked: we experience our cities not only by performing activities at specific destinations, but also by moving from one place to another.
  In this paper, we propose to analyze and cluster the city based on how people move \emph{through} it. Particularly, we introduce \emph{Mobilicities}, automatically generated travel patterns inferred from mobile phone network data using NMF, a matrix factorization model.
  We evaluate our method in a large city and we find that \emph{mobilicities} reveal latent but at the  same time interpretable mobility structures of the city. Our results provide evidence on how clustering and visualization of aggregated phone logs could be used in planning systems to interactively analyze city structure and population activity.
\end{abstract}


\section{Introduction}\label{introduction}

The increasing availability of digital footprints, such as Web/App
access logs, user-generated content, and mobile phone network data, has
allowed to characterize the city at spatio-temporal granularities never
seen before. This means that the different functional areas of the city
can be estimated, based not only on how planners thought that the city
would be lived, but on how people actually used the different spaces
available to them \cite{cranshaw2012livehoods}. However, this is not
enough to understand the city. As Charles Montgomery says in his book,
Happy City: \emph{``When we talk about cities, we usually end up talking
about how various places look and perhaps how it feels to be there in
those places. But to stop there misses half the story, because they way
we experience most parts of cities is at velocity: we glide past on the
way to somewhere else. City life is as much about moving \emph{through}
landscapes as it is about being \emph{in} them''}
\cite{montgomery2013happy}.

Since people may spend a considerable amount of time while moving
through the city, and the quality of that time has a strong influence on
mood, health, and productivity \cite{ruger2017does}, it is important to
understand city structure with respect to mobility. Given that the
growth of cities is faster than the capability of traditional methods to
understand the city, it is important to have cost-effective ways to
analyze the city at scale \cite{zheng2014urban}.

In this paper, inspired by Montogomery's ideas, we estimated a
characterization of the city defined by the collective experience of its
several areas. Particularly, we analyzed intra-city transportation
inferred from mobile phone network records, which we represented in a
\emph{Waypoints Matrix}. This matrix, similar to document-term matrices
used in Information Retrieval, was decomposed using Non-Negative Matrix
Factorization (NMF)~\cite{cichocki2009fast}. We interpreted and labeled
the obtained components, which we denoted \emph{Mobilicities}. We
evaluated our pipeline by performing a case study in Santiago, Chile,
using mobile phone network data from the biggest telecommunications
company in the country. Our pipeline delivered interpretable results, in
contrast with a well-established method. We concluded that
\emph{mobilicities} can be used within an intelligent user interface
aimed at mobility and transportation-analysis tasks.

\section{Background and Related Work}\label{background-and-related-work}

There has been a flurry of research of mobile phone network data known
as \emph{eXtended and Call Detail Records} (X/CDR), as evidenced in
recent surveys on the area
\cite{blondel2015survey, calabrese2015urban}. Some examples include:
understanding socio-economic factors on the population
\cite{soto2011prediction}, understanding family and social relations
\cite{david2016communication}, characterizing response to emergencies
and critical events~\cite{moumni2013characterizing}, crime detection
\cite{bogomolov2015moves}, credit scoring \cite{san2015mobiscore},
test of urban theories~\cite{de2016death}; and the provision of a
cost-effective way of understanding population dynamics and behavior in
developing countries\emph{~}\cite{hilbert2016big}.

The mobile phone network events of a given device depict a
spatio-temporal trajectory that can be processed to infer trips, by
using geometric approaches based on transportation rules
\cite{graells2016day}, or by clustering events in the trajectory
\cite{calabrese2011estimating}. When individual trips are known, it is
possible to aggregate them into Origin-Destination (OD) matrices. This
analysis is common in the literature from X/CDR
\cite{alexander2015origin, iqbal2014development, frias2012estimation, calabrese2011estimating},
and shows that inferring individual mobility is a relevant problem.

Other important aspects are the characterization of land use
(\emph{e.g.}, residential areas, business areas, \emph{etc.}) and
functional areas (\emph{i.e.}, delimited areas that serve specific or
multiple land uses). Since it is crucial to understand the dynamics of
these aspects, functional areas have been measured, monitored and
categorized using digital footprints
\cite{yuan2012discovering, vaca2015taxonomy} and X/CDR
\cite{noulas2013exploiting, lenormand2015comparing, toole2012inferring, ahas2010daily, graells2016sensing}.
A similar work to ours has applied NMF to understand trip purpose, and
build functional areas based on the spatial distribution of such
purposes \cite{peng2012collective}.

The key difference between the aforementioned work and our proposal is
the focus. Other work focuses on the destinations of trips, as well as
activities performed \emph{within} places. As such, their definition of
functional area is limited by those places that, in transportation
terms, \emph{attract} people \cite{hall2012handbook}; yet, as mentioned
by urbanists, the city is experienced in sequence by moving from one
place to another \cite{montgomery2013happy}. Each citizen has an unique
version of the city, built upon the sequence of nodes, landmarks, and
paths traversed \cite{lynch1960image}. In this paper we show that, by
using mobility inferred from X/CDR, and using NMF to decompose/cluster
the different cities experienced by mobile phone users, we are able to
identify the different \emph{Mobilicities} that comprise a big urban
city.
Even though we have centered the discussion around mobile phone network
data, it is possible to infer transportation and urban patterns from
other sources, particularly social media. Twitter has been shown to
be a good predictor of commuter flows \cite{mcneill2017estimating} at
several scales \cite{liu2015multi}. Now, these approaches have the same
limitation as previous approaches: a focus on the origins and
destinations of trips, mainly due to their way of modelling mobility:
using gravity and radiation models (see \cite{masucci2013gravity} for a
comparison). Twitter data, while massive and
longitudinal, does not allow to infer within-trip behavior.

\section{Methods}\label{methods}

Our methods can be summarized in a pipeline of three steps:

\begin{enumerate}
\def\labelenumi{\arabic{enumi}.}
\tightlist
\item
  \emph{Trip detection} from X/CDR data, which, for each device in the
  dataset, identifies its corresponding daily trips, with origin,
  intermediate, and destination towers.
\item
  The construction of a \emph{Waypoint Matrix} \(W\) that aggregates the
  intermediate towers of trips, of a given period of time, into a
  device-antenna matrix.
\item
  The decomposition of \(W\) using NMF into the product of two matrices,
  \(U\) and \(T\), according to a number of \(k\) components, which we
  denote as \emph{mobilicities}.
\end{enumerate}

\textbf{Trip Detection}. To detect trips from X/CDR traces, we resort to
an algorithm based on transportation rules and trajectory simplification
\cite{graells2016day}. The algorithm builds a space-time trajectory
from daily X/CDR events, where space is the cumulative distance between
consecutive connected towers, starting from zero at the first connection
of the day. This trajectory is simplified using a line-simplification
algorithm. Then, each segment from the simplified trajectory is
categorized according to transportation rules, such as the relationship
between the approximated trip distance and time, which is visually
inspected through the slope of the segment. In other words, the trip
detection allows us to separate X/CDR events into the following:
\emph{stationary} events (the user was performing an activity),
\emph{trip start} events (denoting the origin), \emph{trip end} events
(denoting the destination), and \emph{within-trip} events (denoting
mobility).

\textbf{Building the Waypoints Matrix}. Ideally, characterization of
within-trips events does not need to be done using aggregation. For
instance, GPS data allows to do rich clustering over specific
trajectories \cite{zheng2009mining}. However, due to the billing
purpose of X/CDR data, it is possible that trips have few within-trip
events because of the billing cycle. Since we will focus on within-trip
events, we need to aggregate such events from an extended period of
time. Furthermore, some trips do not have within-trip events, such as
those with duration near to the billing cycle time, and those within
zones with low tower density. Hence, by aggregating all within-trip
events for a user in a period of time, the likelihood of identifying the
towers that characterize a specific user's mobility increases. We use
this schema to define a \emph{Waypoints Matrix} \(W\), defined as: \[
w_{i,j} = \frac{\# \text{ of \emph{within-trip} events of user } u_i \text{ at tower } t_j}{\# \text{ of \emph{within-trip} events of user } u_i}
\] This schema is equivalent to the L1-normalized document-term matrices
found in Information Retrieval, but without weighting with Tf-Idf
\cite{yates2011modern}. We do not apply Tf-Idf because its purpose is
to identify discriminative features; conversely, we want to extract
collective features. Additionally, note that this matrix is different to
those used in related work with NMF decompositions
\cite{peng2012collective}: there is a semantic difference between
within-trips and trip start/end events. To avoid this polysemic
behavior, we focus only on within-trips events.

\textbf{Applying Non-Negative Matrix Factorization}. To represent how users interact
with towers, we propose to decompose this matrix into two:
\(W = U \times T\), where U is a \(|u|\times k\) matrix that encodes
\(k\) user latent features for \(|u|\) users, and \(T\) is a
\(k \times |t|\) matrix that encode \(k\) latent tower features for
\(|t|\) different cell phone towers. Note that, by definition, all
\(w_{i,j} \geq 0\). NMF allows us to decompose the matrix W into two
non-negative matrices, which gives a lower rank approximation for \(W\),
such that \(W \approx U \times T\) \cite{kuang2012nmf}. This can be
formalized as the following optimization problem:
\(\min_{U,T} \|W - U \times T\|_F\) subject to \(U\) and \(T\) be
non-negative, where number of rows in \(U\) and the number of columns in
\(T\) correspond to the desired lower-rank approximation \(k\).

Even though there is a variety of methods to decompose matrices, we
choose NMF, which has been applied in similar contexts with
interpretable results \cite{peng2012collective}. Then, we define a
mobilicity \(m_c\) as the weighted set of towers within the \(c\)
component of the decomposition, \emph{i.e.}, the \(c\)-th column of
matrix T. The parameter \(k\) must be chosen manually, and its value
should be ideally decided jointly between data scientists and domain
experts according to the context. Previous strategies for choosing
\(k\) have focused in measuring the stability of the components
\cite{brunet2004nmf} and in the variation of the residual sum of
squares curve between the original matrix and its decomposition
\cite{frigyesi2008nmf, hutchins2008nmf}. However, we prefer to manually
choose the number of components as these methods do not allow us to
incorporate external information such as the socio-economic distribution
of the city.

\section{Case Study: Santiago, Chile}\label{case-study-santiago-chile}

\begin{figure}[t]
\centering
\includegraphics[width=0.9\linewidth]{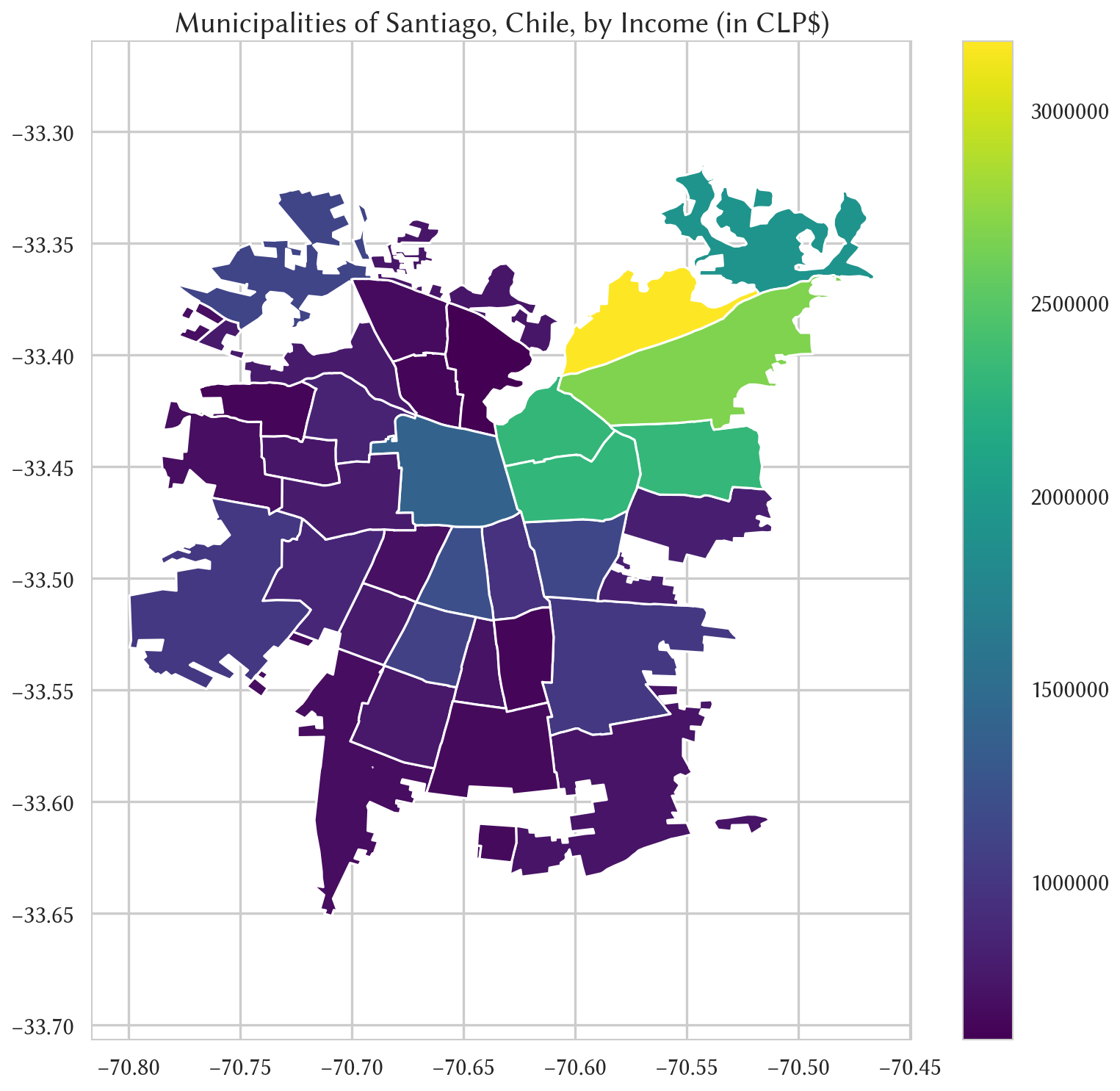}
\caption{Choropleth map of the urban area of Santiago, Chile. Each municipality is colored according to their average income (in CLP\$). \label{fig:stgo_income}}
\vspace{-4mm}
\end{figure}

We performed a case study on Santiago, the capital of Chile, with almost
8 million inhabitants. Its urban area covers a surface of 867.75 square
kilometers, and is composed of 35 independent administrative units
called municipalities (\emph{c.f.} Fig.~1). Because this city has experienced accelerated
growth, and it is expected to keep growing at least until 2045
\cite{puertas2014assessing}, understanding its structure at scale is an
important and timely task.

\subsection{Datasets}

\begin{figure}[t]
\centering
\includegraphics[width=0.9\linewidth]{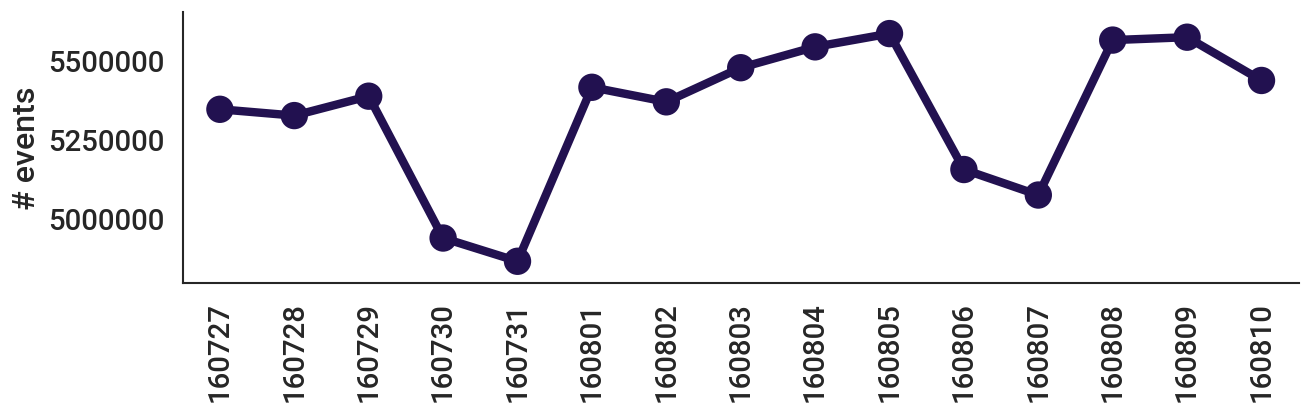}
\caption{Number of events in the dataset per day. The effect of weekends in the number of events is easily identifiable. \label{fig:event_count}}
\end{figure}

\textbf{Mobile Phone Network Data}.
We studied an anonymized X/CDR dataset from Telefónica Chile, the
biggest telco. in Chile, with a market share of 33\% in 2016. The
dataset contained records between July 27th and August 10th from 2016.
In total, we analyzed 124,414 users, who had enough connections to the
cell towers under analysis to estimate their daily trips between 6AM and
midnight, and had either a pre-paid or contract subscription. They
generated an average of 5.33 million billing records per day
(\emph{c.f.} Fig. \ref{fig:event_count}). The average inter-event times
for users range within 14.71 and 30.96 minutes, which shows a billing
cycle between fifteen and thirty minutes.

\begin{figure*}[t]
\centering
\includegraphics[width=0.9\linewidth]{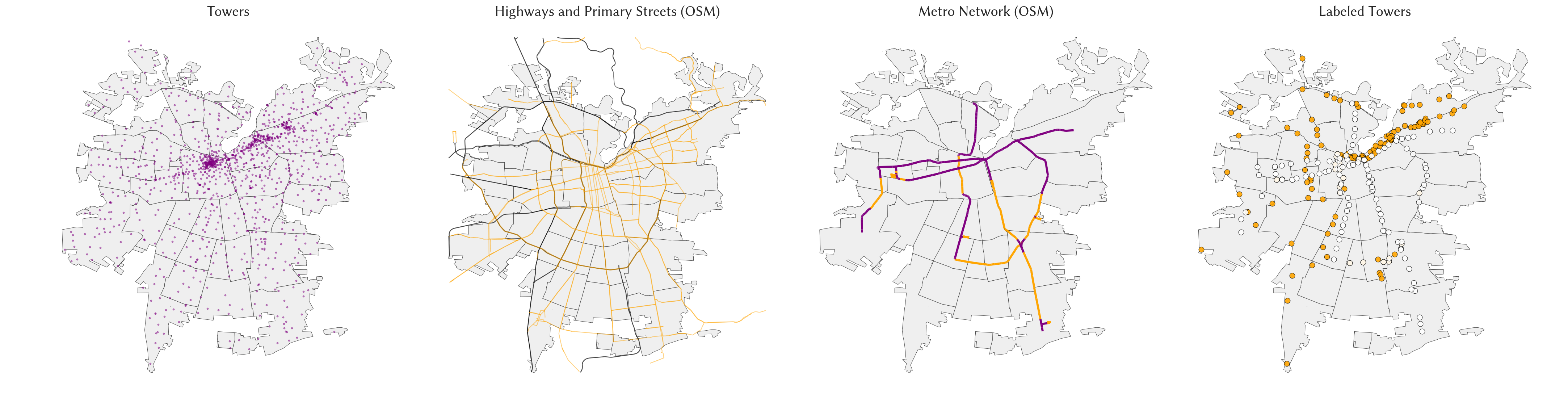}
\caption{Maps of Santiago: cell phone tower network, highway and primary streets, the metro network, and the set of labeled towers according to their distance to highways or metro lines. \label{fig:santiago_maps}}
\end{figure*}

Telefónica has 1,464 cell phone towers in
the municipalities under consideration. We discarded towers that were
installed in in-door contexts (\emph{e.g.}, malls, hospitals,
\emph{etc.}). This is possible because tower meta-data includes their
geographical position and their name. For out-door towers, the name
usually contains the nearest crossing, while in-door towers contain the
name of the place they lie in. In total, there were \(|t| =\) 1,082
out-door towers (see Fig. \ref{fig:santiago_maps}, Towers). The only
in-door towers that we kept were those installed within underground
metro stations.

\textbf{OpenStreetMap}. OSM (\url{http://openstreetmap.org}) is a
crowdsourced maps platform. We downloaded a dump of its data for Chile,
and then identified the highways within Santiago. We used
this information to contextualize the different \emph{mobilicities}
identified by the NMF. We did so by finding the out-door towers that lie
within 250 meters of each highway, as shown in Fig.
\ref{fig:santiago_maps} (Labeled Towers).

\subsection{Trip Detection and Waypoints Matrix}

\begin{figure}[t]
\centering
\includegraphics[width=0.95\linewidth]{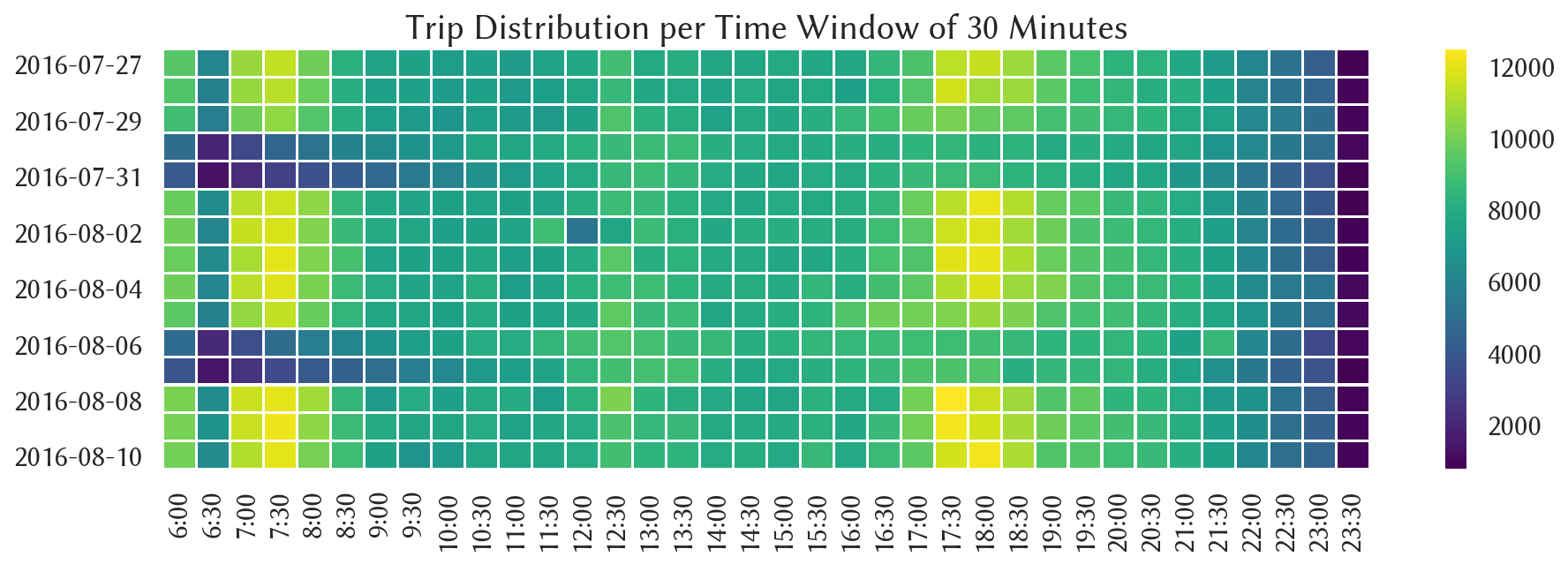}
\caption{Trip departure time distribution per day.\label{fig:trip_start_time}}
\vspace{-4mm}
\end{figure}

Using the trip inference algorithm we detected 4,213,400 trips for 124,415 users (\textit{c.f.} Table \ref{table:trip_stats} for descriptive statistics). 
Fig.~\ref{fig:trip_start_time} shows the departure time distribution of all trips. One can see that business days exhibit expected peak-times related to work hours, and that weekends exhibit different patterns, such as a higher density at lunch time.

\begin{table}[t]
\centering
\begin{tabular}{lr}
\toprule
{Metric} &               \\
\midrule
\# of Trips & 4,213,400 \\
\# of Users & 124,415 \\
\# of Users with Within-Trip events & 95,027 \\
\midrule
Mean trips per user  &      33.87 \\
Std. Dev.   &      19.62 \\
Min   &       1 \\
Percentile 25\%   &      18 \\
Percentile 50\%   &      34 \\
Percentile 75\%   &      48 \\
Max   &     140 \\
\bottomrule
\end{tabular}
\caption{Statistics with respect to the number of inferred trips.}
\label{table:trip_stats}
\end{table}

After detecting trips, we built the \(W\) matrix, of dimensions
\(|u| \times |t|\). Note that \(|u| =\) 95,027, because not all users
had within-trip events. 

\begin{figure*}[t]
\centering
\includegraphics[width=0.9\linewidth]{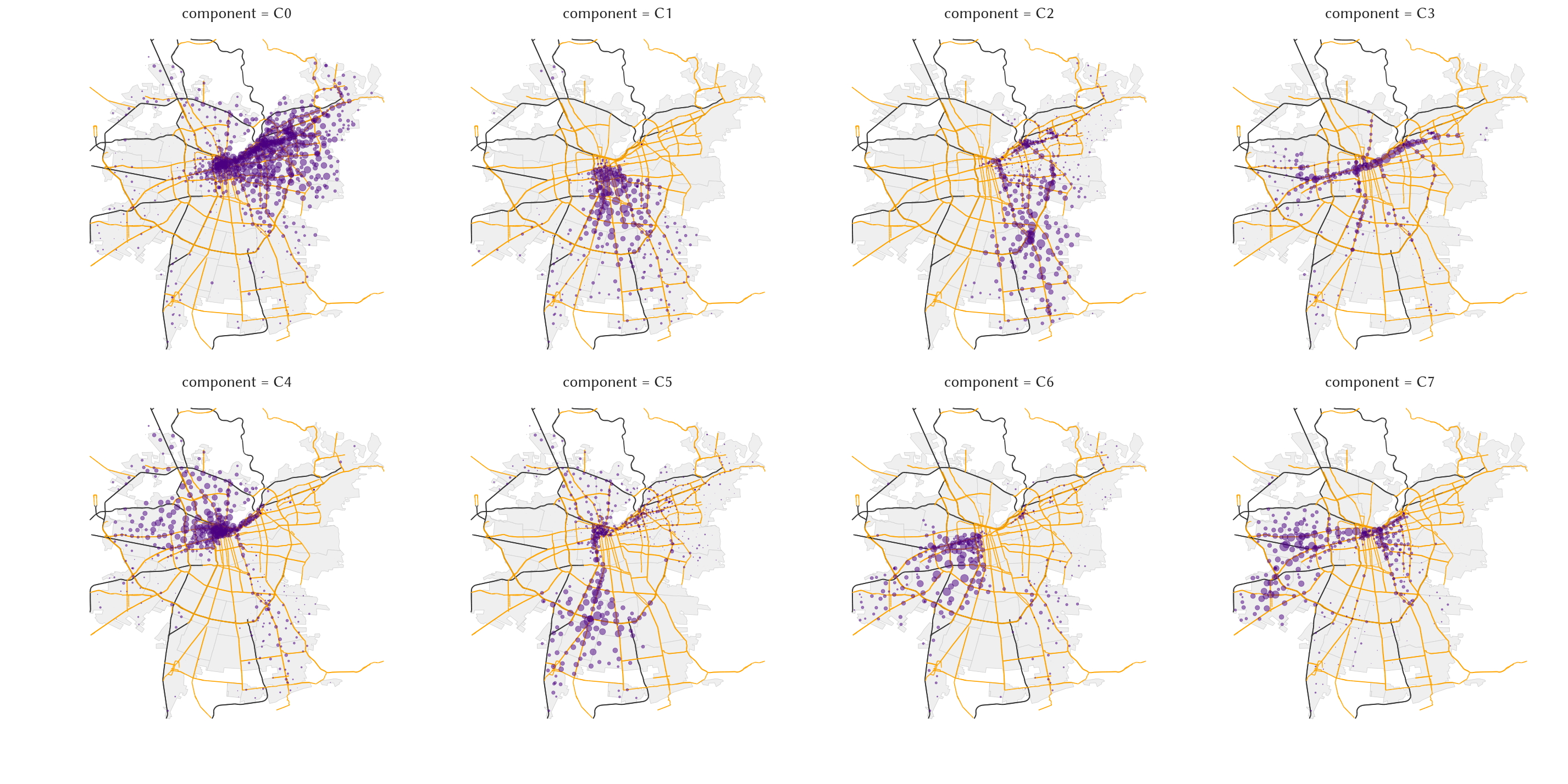}
\caption{The eight \emph{mobilicities} of the tower-component matrix obtained by performing NMF. \label{fig:nmf_components}}
\end{figure*}

\subsection{Factorization of the Waypoints Matrix}
To perform the NMF decomposition, we chose
\(k =\)~8, because the city is usually divided into six big areas
(north, south, east, west, south-east, center) and, since we expect that
the results exhibit relationship with modes of transportation, we wanted
to see the effect of private and public transportation. Thus, \(k =\) 8
is an arguably reasonable choice (note that we discuss the choice of $k$ at the next section).

\textbf{Tower-Component Matrix}.
Figure \ref{fig:nmf_components} shows the results of the factorization, with one map for each component-tower column from the matrix. One can see that there is a strong geographical clustering of towers, which may be explained as \(W\) is essentially a co-occurrence matrix.

\begin{figure*}[t]
\centering
\includegraphics[width=0.9\linewidth]{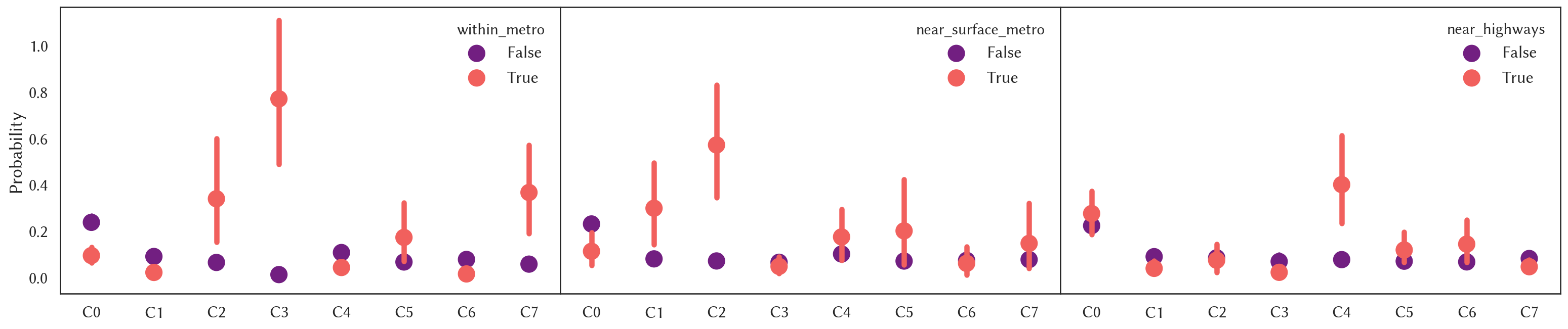}
\caption{Point-plot of the average association of the labeled sets of towers into the different \textit{mobilicities}. \label{fig:nmf_tower_factors}}
\vspace{-2mm}
\end{figure*}

Fig. \ref{fig:nmf_tower_factors} show how the sets of labeled towers (those near highways, near surface metro and within underground metro stations) relate to each component. This allows to see that some components tend to be more associated than others to some modes of transportation: C1, C2, C3 and C7 are more associated to metro than highways, while C4 exhibits the opposite behavior. Having both figures into account, the following is an interpretation of each \emph{mobilicity}:

\begin{itemize}
\tightlist
\item
  [C0]: the east side of the city, including part of the center, next to
  the yellow metro line and one important highway of the city. This area
  is characterized for its business districts and high income
  residential areas (\textit{c.f.} Fig. \ref{fig:santiago_maps}). As such, it is likely that its residents do not use
  public transportation, nor visit other \emph{mobilicities}. Note how metro towers have lower association with this component in Fig. \ref{fig:nmf_tower_factors}.
\item
  [C1]: people that live in the southern part of the city, mostly between
  two metro lines. Since this area is characterized by low income, this means that they need to take
  a bus to reach the metro. 
\item
  [C2]: the southeast area of the city, which is characterized by their
  dependency of two metro lines. This component contains mixed-income municipalities.
\item
  [C3]: In contrast with the previous components, this one is completely focused on public transportation: it fully contains two metro lines in full, and partially other two. It also contains bus corridors that tend to connect to metro lines. 
\item
  [C4]: the northern part of the city, which is mostly residential and of low income. The component also has a main street of the city as a kind of tentacle, showing that people who lives/work in this area, but who work/lives in another, uses this street as a way to get into the component.

\item
  [C5]: the south-west part of the city, which is connected to downtown primarily
  through a highway and a metro line that is parallel to the highway.

\item
  [C6]: the western area of the city. This area contains one of the most populated municipalities in the city.
\item
  [C7]: similar to C6, but extending its reach to center areas of the city through a metro line a bus corridors. This makes this component dependent on public transport, and thus, the routes followed by its inhabitants tend to cluster, in contrast to what happens in C6.
\end{itemize}

In summary, latent cities seem to be comprised by three kinds of clusters of towers: those where people lives and moves, enclosed by specific limits (C0, C1, C6), those where people lives and work, but in different areas of the city, connected through the transportation network (C2, C4, C5, C7), and transportation infrastructure (C3). 

\begin{figure}[t]
\centering
\includegraphics[width=0.95\linewidth]{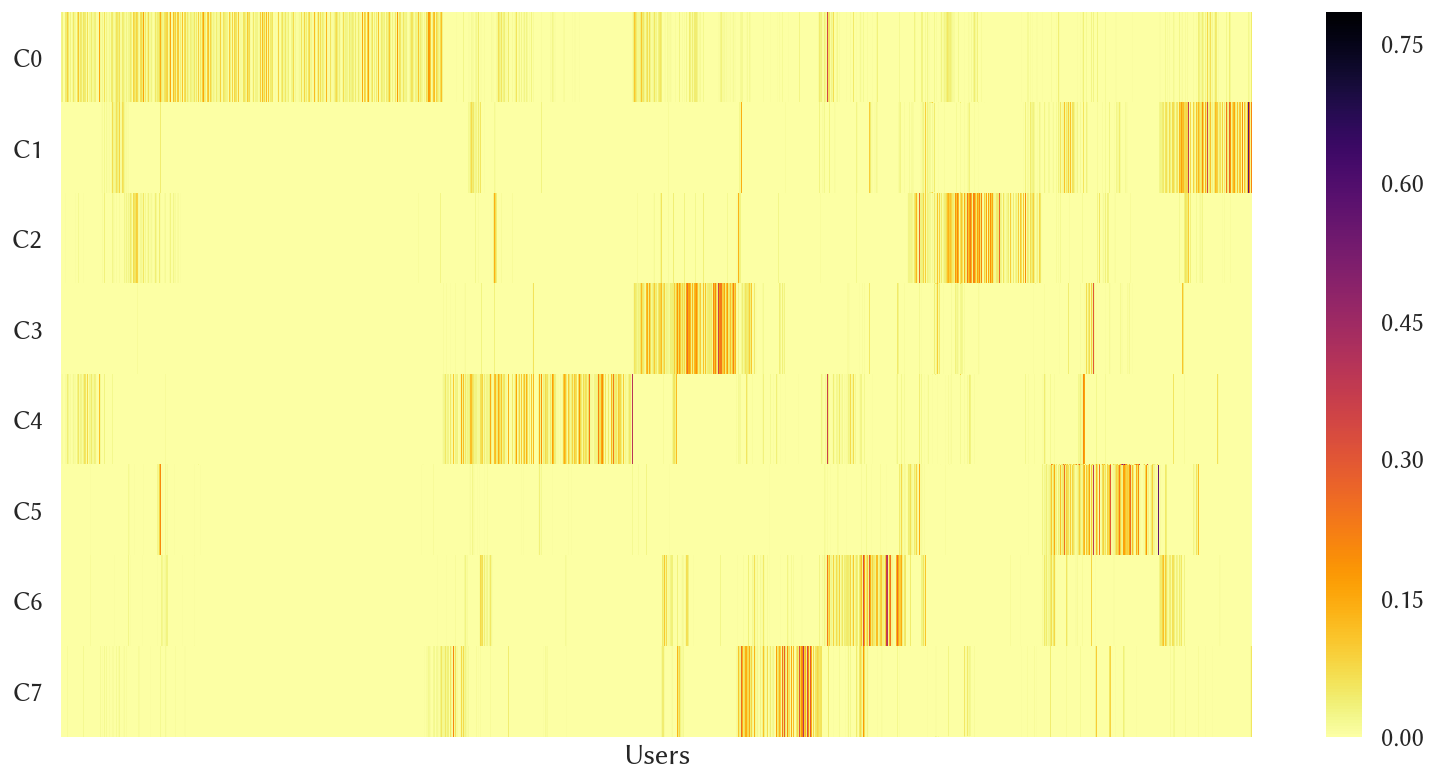}
\caption{Heatmap of a random sample of 25K users and their corresponding component associations. \label{fig:nmf_users}}
\vspace{-4mm}
\end{figure}

\textbf{User-Component Matrix}.
The user-component matrix may suggest that users pass through different \emph{mobilicities} in their daily lives.  Fig. \ref{fig:nmf_users} explores this potential behavior, by displaying how a sample of 25,000 users cluster around the corresponding components. One can see that, indeed, users tend to have a primary component, but they still belong to others. This would be the equivalent to, for instance, living in the suburbs, and having to travel long distances to go to work. Note that many users from C0, the wealthiest part of the city, have negligible association to other components -- something expected due to the economical segregation of the city (\textit{c.f.} Fig. \ref{fig:santiago_maps}).

\begin{figure}[t]
\centering
\includegraphics[width=0.9\linewidth]{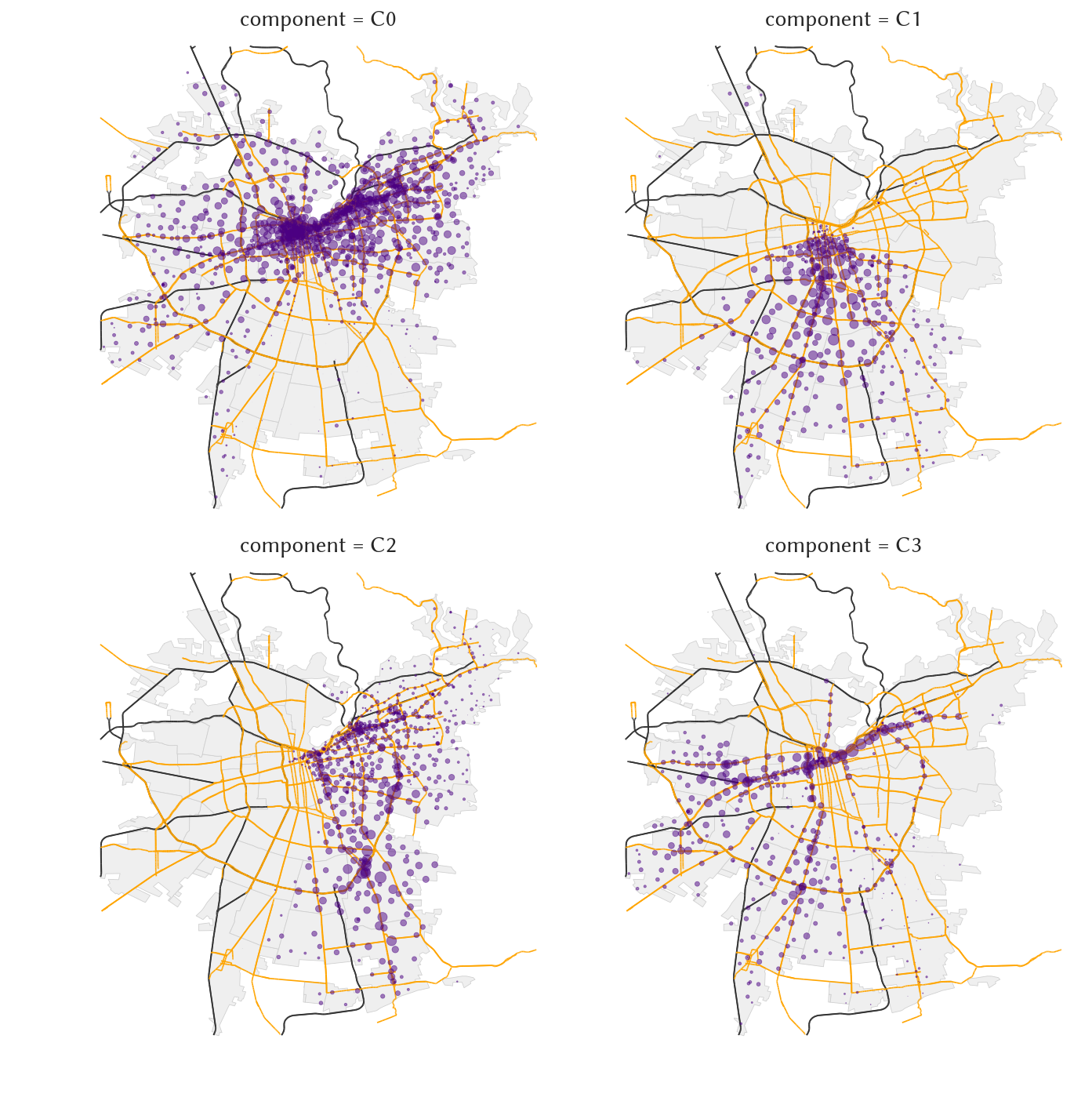}
\caption{Mobilicities obtained with $k = $4. \label{fig:nmf_components_k4}}
\vspace{-5mm}
\end{figure}

\begin{figure*}[tp]
\centering
\includegraphics[width=0.9\linewidth]{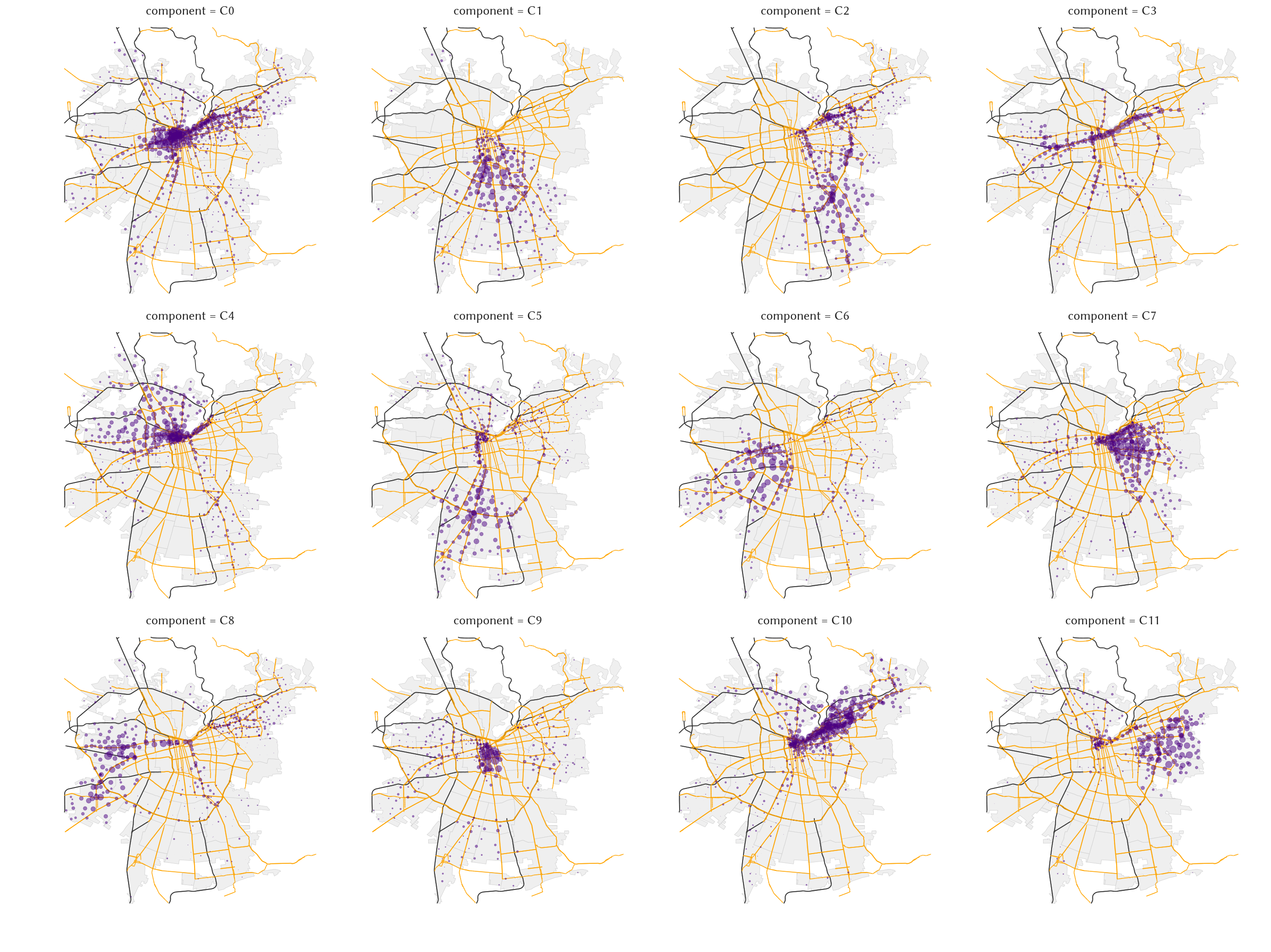}
\caption{Mobilicities obtained with $k =$12. \label{fig:nmf_components_k12}}
\vspace{-5mm}
\end{figure*}

\textbf{Understanding $k$}.
We showed the eight \emph{mobilicities} (\emph{c.f.} Fig.~\ref{fig:nmf_components}) to domain experts, who gave informal feedback -- it made sense to split the city in this way, as we have interpreted earlier. Even though a formal evaluation with domain experts is left for future work, here we discuss the patterns that emerge when varying the parameter $k$ which is the rank of the factorized matrix. We do so by estimating mobilicities with $k' = $ 4 (\textit{c.f.} Fig. \ref{fig:nmf_components_k4}) and $k'' = $ 12 (\textit{c.f.} Fig. \ref{fig:nmf_components_k12}).

With four mobilicities, the clustering is mostly geographic: three components split the city. However, the fourth component is related to public transportation: it reconstructs several metro lines and bus corridors. In this aspect, it seems that using $k' =$ 4 allows to obtain a similar result to $k = $ 8.  Then, if one would like to differentiate cell towers with respect to general transportation patterns, this could be a reasonable choice.

With twelve mobilicities, the geographical clustering is still present, but the routes that connect distinct parts of the city become more evident -- meaning that a mobilicity is comprised by one or two sectors of close towers (for instance, home and work locations), plus ``bridges'' that connect one mobilicity to another, based on the common routes followed by people. This behavior is expected, due to the co-occurrence property of the Waypoints Matrix. 

In summary, several values of $k$ allow to infer soft-partitions of the city, as well as the way its inhabitants move between those partitions. A mobilicity may be a soft-partition, a soft-partition with bridges to other mobilicities, or a network of those bridges -- namely, a transportation network.

\begin{figure*}[tp]
\centering
\includegraphics[width=0.9\linewidth]{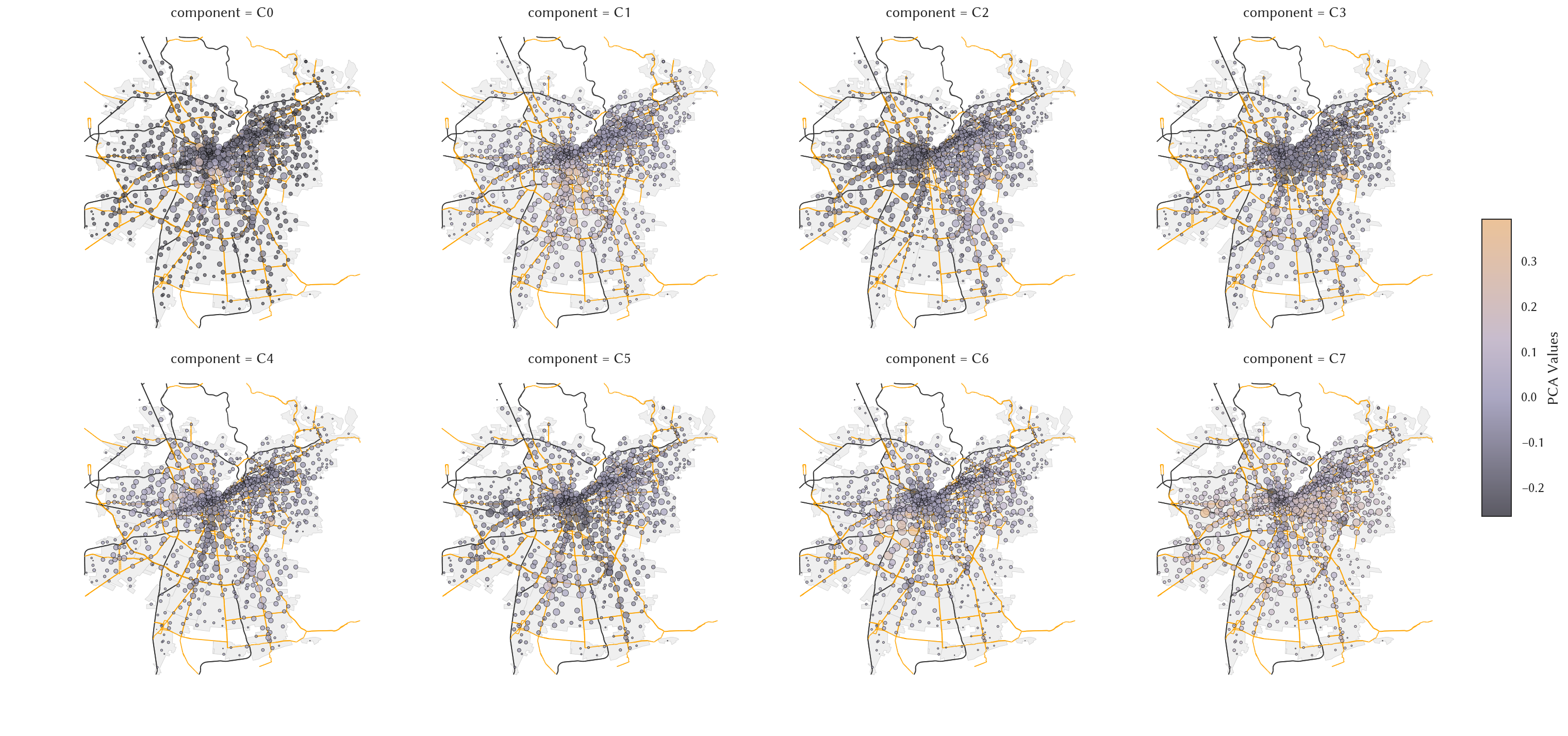}
\caption{Tower-component results of the application of Truncated SVD/PCA to the Waypoints Matrix, with $k =$ 8.\label{fig:pca_components}}
\end{figure*}

\textbf{Comparing Interpretability with PCA/Truncated SVD}.
To discuss further whether NMF is a good choice of model in terms of interpretability, we
estimated a Truncated SVD decomposition (equivalent to PCA) with \(k =\) 8 . Fig.
\ref{fig:pca_components} shows that, in contrast to NMF, there is no
geographical clustering nor correspondence to any infrastructure
available in the city. Thus, even though PCA is a widely used dimensionality reduction technique, it does not allow the interpretation nor clustering of the city in the same way as NMF does. 

\section{Conclusions}

We proposed the concept of \emph{mobilicities}, which denotes the
different cities experienced by the inhabitants of a big city, and
depict its dynamics with respect to mobility and usage of modes of
transportation. The suitability of NMF to this kind of spatial data could be related to the fact that NMF is equivalent to spectral clustering \cite{ding2005equivalence}, which has performed well when grouping trip destination data \cite{cranshaw2012livehoods}. However, as we have noted in our motivation, our input is not destination nor origin data; instead, it is \emph{spatial location while moving}. This focus was inspired by the book Happy City \cite{montgomery2013happy}, reflecting
that our purpose was to help domain experts and policy designers to make
better, happier cities. Such purpose implies collaboration between the emerging
field of data science and the corresponding disciplines -- transportation
and urban planning. However, evidence-based policy in those areas requires transparency
and interpretability, and many state of the art machine learning
techniques do not offer both qualities \cite{castelvecchi2016can}. In
this paper, we have shown that NMF does offer both qualities when
applied to mobility data, and thus, is a promising technique to apply in
the field of Urban Computing \cite{zheng2014urban}.  


\textbf{Limitations and Future Work}.
Critics may rightly say that we need a well-defined criteria to choose \(k\). Future work should tackle this limitation using intelligent user interfaces aimed at domain experts. This opens two lines of research within the IUI: on the one hand, we could try other factorization methods for positive-only data such as SLIM, which has shown promising results in the past~\cite{Larrain2015}, and would allow to understand how the choice of $k$ influences the output and its interpretability. 
On the other hand, visualization and exploratory interfaces are tools
valued by domain experts~\cite{cheng2013exploratory}, and mobility has
been a recurring topic in visual analytics~\cite{andrienko2013visual}.
Finally, our work did not consider the temporal aspects of transportation. Hence, future work should consider how to incorporate that dimension into the definition of Mobilicities.

\textbf{Acknowledgements}.
The analysis was performed using Jupyter Notebooks \cite{perez2007ipython}, jointly with the \emph{scikit-learn} \cite{scikit-learn}, \emph{pandas} \cite{mckinney2010data}, and \emph{geopandas} libraries. The maps on this paper include data from \textcopyright OpenStreetMap contributors. We also thank Telef\'onica R\&D in Santiago for facilitating the data for this study, in particular Pablo García Briosso. 
The author Denis Parra has been funded by Conicyt, Fondecyt grant 11150783, as well as Fondef grant id16i10222 and the BRT+ Centre of Excellence funded by VREF.
Finally, we thank the anonymous reviewers for the insightful comments that helped to improve this paper.

\small
\bibliography{references}
\bibliographystyle{plain}

\end{document}